# Polyaniline as a conductive polymer and its role in improving the efficiency and conductivity of perovskite solar cells

**Vahdat Nazerian[1], Mehran Hosseinzadeh Dizaj[2], Tole Sutikno[3,4]**
[1]Department of Electrical Engineering, University of Mazandaran, Babolsar, Iran
[2]Department of Electrical Engineering, Central Tehran Branch, Islamic Azad University, Tehran, Iran
[3]Department of Electrical Engineering, Faculty of Industrial Technology, Universitas Ahmad Dahlan, Yogyakarta, Indonesia
[4]Embedded System and Power Electronics Research Group, Yogyakarta, Indonesia

**ABSTRACT**

This article investigates the role of polyaniline as a conductive polymer in the active layer of perovskite solar cells. Samples were created by incorporating polyaniline into the transport layers to assess its impact on enhancing efficiency and conductivity. The application of this polymer across various layers of the cell structure led to improved stability and performance. Given its high doping capability, polyaniline was examined in detail, particularly focusing on two types of oxidation doping and its integration into the hole transport layer. Graphene oxide and reduced graphene oxide were chosen as comparative models, and their performance was evaluated against the standard polyaniline configuration. Laboratory results revealed that power conversion efficiency increased by 17.5% with graphene oxide and by 36.8% with reduced graphene oxide. Furthermore, short-circuit current density improved by 9.8% and 23.1%, respectively. These findings are consistent with existing studies in the field and support the validity of the approach.

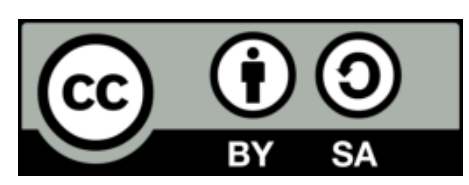

***Corresponding Author:***

Vahdat Nazerian
Department of Electrical and Computer Engineering, University of Mazandaran
Pasdaran Street, Babolsar, Iran
Email: v.nazerian@umz.ac.ir

## 1. INTRODUCTION

Polymers are very large molecules that are formed by connecting several hundred monomers to each other and forming long chains. The word "polymer" is of Greek origin and is composed of the word "poly," meaning "several," and "mer," meaning "part." A monomer is the smallest repeating unit of a polymer. Among conductive polymers, polyaniline (PANI) has attracted much attention due to its chemical stability, high conductivity, and electrochemical properties. Recently, PANI has received attention due to its applications in electrical devices from the point of view of research and industry. But PANI has low solubility in organic solvents, is insoluble, and is therefore almost un-processable. Today, PANIs are used in solar cells, especially perovskite solar cells (PSCs) [1], [2].

The electrical conductivity and intrinsic polymer characteristics play a crucial role in enhancing the performance of perovskite solar cell layers. Conductive polymers containing conjugated π-electron systems, such as C=C bonds, exhibit unique electronic properties and demonstrate greater ease of oxidation and reduction compared to conventional polymers. Materials like polypyrrole, polythiophene, and polyaniline possess complex dynamic structures that are widely utilized in the development of smart materials. When subjected to electrical stimuli, these polymers undergo significant changes in their chemical, mechanical, and electrical properties. A comprehensive understanding of their synthesis and the extent to which their properties can be modulated by electrical input is essential for achieving controlled functionality in advanced applications [3], [4].

In the context of solar cell fabrication, the incorporation of conductive polymers contributes to improved electrical characteristics, including enhanced conductivity and electron mobility within the active layers. Specifically, the use of polyaniline (PANI) in both the hole transport layer (HTL) and electron transport layer (ETL) has been shown to increase electron mobility by approximately 10% [5]. These enhancements support more efficient charge transport and overall device performance, aligning with findings from previous studies [3], [4]. Such results underscore the potential of conductive polymers as key components in next-generation photovoltaic technologies.

## 2. METHOD

The presentation picture of the experimental setup (such as the spin coater device, the isolated chamber, and the sputtering coating machine) and the advanced solar simulator are shown and described in detail in Figure 1. In general, the work of spin coating involves applying a thin layer (from nanometers to micrometers) uniformly across the surface of the intended substrate by coating a solution of the desired material in a solvent while it is rotating. Simply put, a liquid solution is placed on a rotating bed to produce a thin layer of solid material, such as a polymer, as shown in Figure 1(a). Aniline (2 cc) is first poured into an Erlenmeyer flask and placed in an ice bath, as shown in the spin coating setup in Figure 1(a). While the spin coater is operating, 30 cc of 1 M hydrochloric acid is gradually added using a pipette to form solution A. In a separate flask, 4 grams of ammonium persulfate are mixed with 20 cc of 1 M hydrochloric acid to produce solution B.

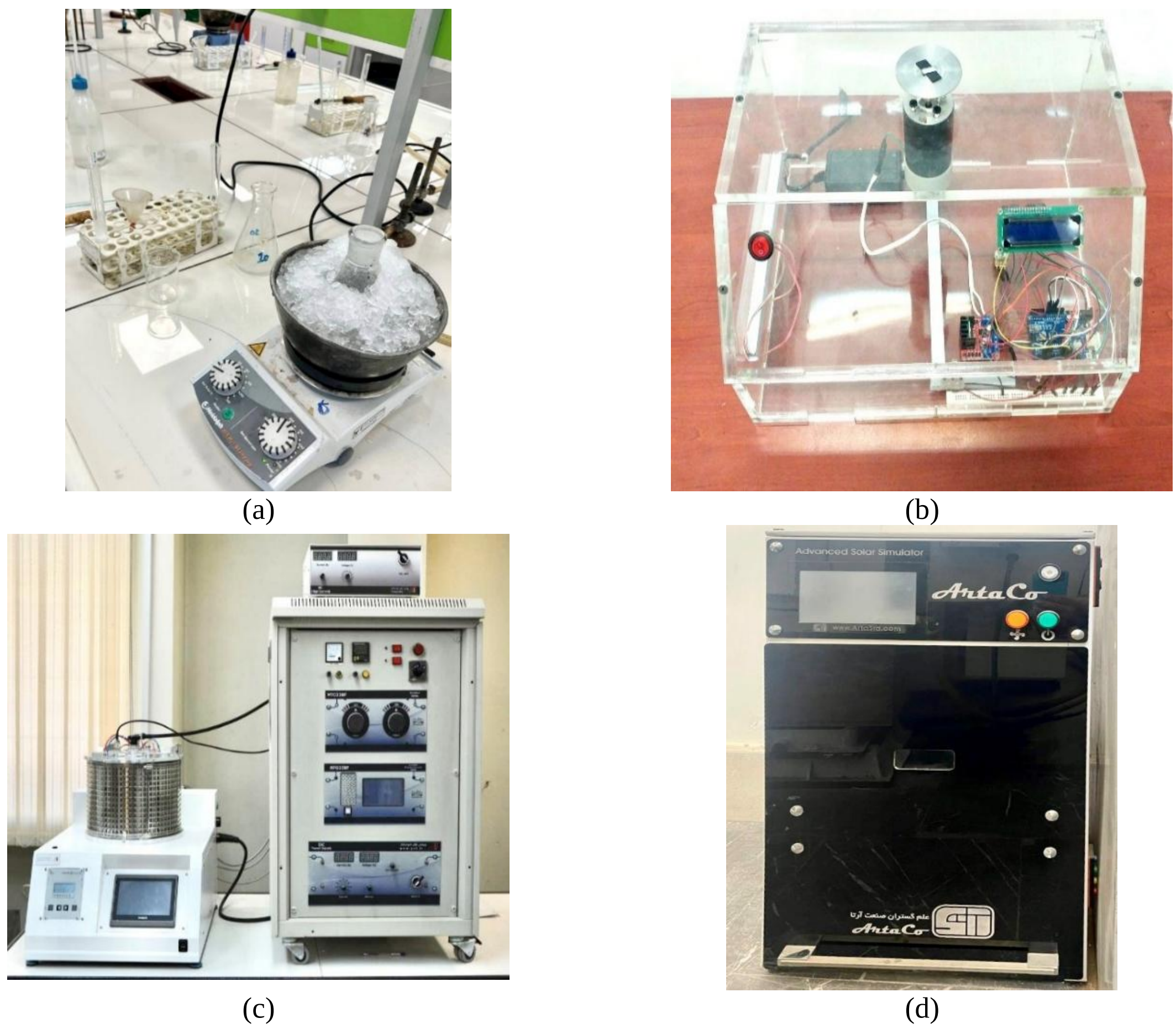

Figure 1. Presentation picture of the experimental setup and the simulator: (a) spin coating with an Erlenmeyer flask in an ice bath placed on the rotating bed, (b) an isolated chamber, like a glove box, for layering the perovskite material, which is very toxic, (c) a sputtering machine for coating of ETL and HTL layers, and (d) an advanced solar simulator to measure the optical and electrical properties of PSCs

Solution B is added to solution A to synthesize polyaniline with over 90% purity. The product is then filtered and dried in an oven for one hour. After preparation, polyaniline is used to make perovskite solar cells. The perovskite layer is applied inside an isolated chamber to safely handle its toxic nature, as shown in Figure 1(b). The ETL and HTL layers are coated using a sputtering machine, as shown in Figure 1(c). Finally, the solar cell is tested with a solar simulator to measure its optical and electrical performance, as shown in Figure 1(d).

After preparing the samples of PANI in transport layers, we used the advanced solar simulator shown in Figure 1(d), to measure the optical and electrical properties of the modules to investigate the role of these layers in the efficiency and conductivity of PSCs. In fact, the advanced solar simulator is a powerful tool for characterizing the performance and efficiency of solar cells, and can provide valuable information for researchers in the solar industry. Also, the standard synthesis method of PANI and the material obtained are shown in Figure 2.

a) Pour 2 cc of aniline monomer into an Erlenmeyer flask and place it in an ice bath (because the process is exothermic).
b) Then add 30 cc of 1 M HCL to it evenly with a pipette.
c) We call the obtained material A.
d) Now pour 4 grams of ammonium persulfate into another Erlenmeyer flask and add 20 cc of 1 M HCL gradually while stirring.
e) We call the obtained second material B.
f) We add substance B drop by drop to substance A. (Be careful, we add B to A), as shown in Figure 2(a).
g) It is stirred for 1 hour in an ice bath and then washed with 1 M acid and filtered.
h) In the last step, we dry it in the oven for 1 hour.
i) The material obtained is called PANI [6]–[8].
j) The obtained material is black in color, as in Figure 2(b).

PANI is one of the artificial conductive polymers, whose high electrical conductivity has received a lot of attraction. Basically, PANI is known as a redox polymer, and it is prepared by chemical and electrochemical methods in an acidic environment. The choice of which method depends on the kind of application. If you need thin films with better properties and purity, the electrochemical method is preferred. PANI is used in various fields such as microelectronics, corrosion coatings, sensors, and electrodes for batteries due to its thermal and radiation stability, conductivity, ease of synthesis, low cost, and diverse structure [9]–[12].

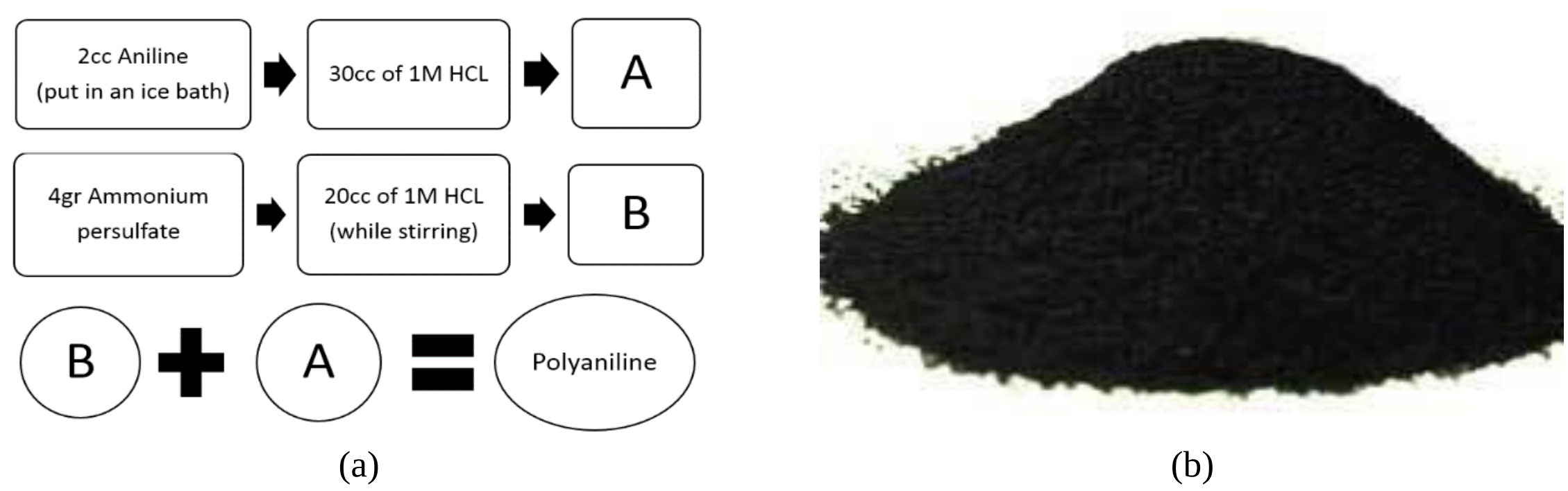

Figure 2. Preparation of polyaniline: (a) standard synthesis method and order of composition of A and B materials, and (b) the state and form of PANI that usually exists

PANI has high conductivity, low toxicity, hydrophilicity, and suitable environmental stability at the level of nanostructures. Cellulosic materials in combination with PANIs give new capabilities and parameters to this structure. PANI networks reinforced with cellulose, shown in Figure 3, are significantly suitable for antibacterial agents, antioxidants, sensors, and electromagnetic shielding devices (electromagnetic field) [13]–[16].

Most of the time, polymers have been used as an insulator due to their lack of electrons and mobile ions. However, some of them make a conductive route when there is an electrical action and become conductive. Bendy named them as a semiconductor. To introduce and express the properties of polymers, we use some common parameters. For example, conductivity, resistance, dielectric constant, capacitor, and loss coefficient [17]–[20]. With all the advantages that we have mentioned, especially in the case of PANI, we will now explain the specific disadvantages that can be seen. For example, being insoluble or slowly soluble in common solvents, polymeric aggregates with long chains, but the most important disadvantage of this, the polymer PANI is that its conductivity decreases during a long cycle. Today, researchers use other methods to reduce these disadvantages, such as copolymerization with PANI derivatives or other polymers, as well as synthesis using functional organic acids [21], [22].

PANIs are one of the most widely used conductive polymers and were discovered in the middle of the 19th century. The polymer chain of PANIs may have units of quinonoid or benzenoid or both at the same time in different amounts. This special conducting polymer can be prepared from aniline monomer through the technique of electrochemical oxidative polymerization or chemical oxidation under acidic conditions. The best and most widely used initiators in chemical methods are potassium persulfate (KPS) and ammonium persulfate (APS). Chemical methods allow the production of this type of polymer composites, as the electrochemical method is suitable for low-scale polymer production. Electrochemical methods include the co-deposition method, as shown in Figure 4 [23]–[26].

Polyaniline (Emeraldine Base)

HCl

Polyaniline (Emeraldine Salt)

Figure 3. Chemical and molecular structure of PANI polymer compound

Chain Initiation

Initiator

Chain Propagation

(a)

Acid

Initiator

Chain Termination

Acid

Polyaniline

(b)

a+b-2H+

Aniline

-2e-, -2H+

Polyaniline

Figure 4. Two types of polymerization: (a) chemical polymerization and (b) electrochemical polyaniline

You can see two types of polymerization above. In Figure 4(a) chemical polymerization is done in an acidic environment using APS and KPS, but in electrochemical polymerization shown in Figure 4(b), it is done in aniline and acid electrolyte solution by applying voltage between the electrode and the counter. The composition of polymer materials plays an important role in their conductivity and stability. PANIs are in 3 oxidation states. It depends on the redox states of the construction as shown in Figure 5.

As explained earlier, excellent materials show different properties according to the placement of them terminate type in polymer chains. In Figure 5, we see different forms of PANI compound chains in different states of (a) PANI, (b) leucoemeraldine, (c) pernigraniline, (d) EDB, and (e) EDS.

Figure 5. Various structures of (a) PANI, (b) leucoemeraldine, (c) pernigraniline, (d) EDB, and (e) EDS

### 2.1. Investigating the properties of polyaniline

Conductive polymers such as PANIs belong to a specific group. Their compositions include several monomer units with conjugated chemical bonds that make electrical conduction possible under certain conditions, such as doping. These polymers can be a good substitute for metals and semiconductors. They are usually used in various industries, such as electronics. Because they have high conductivity, low density, and high flexibility, they can be used, for example, in the process of making flexible PSCs [27]–[29]. The use of conductive polymers such as PANI in the electronics industry is very restricted, because its conductivity at temperatures more than 150 degrees Celsius is unstable due to the discharge of polymer chains, but it can be used as an electrode material in batteries. Among the applications of PANI, we can mention the design of rechargeable batteries. Based on these polymers, a secondary current source was developed on graphite-PANI. These sources have high efficiency compared to analog energy sources, but their main drawback is still their instability. When recharging. As mentioned earlier, PANIs are reduced in their conductivity in a long process, so 0.15% of the energy paste is enriched every time, but their special feature is their ability to be reprocessed and their compatibility with the environment [1], [2], [30], [31].

It is usually said that electrochemical synthesis produces the purest product that does not have any additives and does not require external methods to purify PANI from solvent and monomer, as well as reaction initiator molecules. Regarding the electrochemical synthesis of PANI, the above methods are mentioned as potentiostatic, galvanostatic, and potentiodynamic. Today, PANI-based composite materials are used electrochemically in organic field transistors [32]–[34]. Three types of oxidation composition can be seen according to Figure 6. The effect of pH and also the amount of absorbing monomer of the layers, what effect does it have on the oxidation process.

### 2.2. Polyanilines in perovskite

Now we investigate the optoelectronic properties of PANI compound layers with graphene oxide as a hole transport layer in PSCs in the reverse method [35], [36]. PANI composite films increase the electrical conductivity and improve efficiency compared to the previous intact films. In the case of using the previous layers, the efficiency was about 16.61%, which, after using PANI composites It has increased to 21.6% [37]–[39]. PSCs have reached about 25% by engineers and researchers in the last 5 years by modifying their manufacturing method. ETL is the most proven layer [40], [41].

In the inverse method, due to the process ability and performance even at low temperature, a lot of surface layer materials have attracted very attention as another redesign. In reverse perovskites, the HTL is placed on the transparent conductive oxide anode to make the hole transfer from the perovskite absorber easier, while ETL must exist in a relationship between the perovskite layer and the metal anode [42].

PANI, which is one of the best P-type conductive polymers, is widely used in PSCs today. Inverse perovskites that use HTL based on (PANI) have received much attention today due to their high conductivity, conductive paths, and relatively good environmental and chemical stability. A useful solution for the synthesis of PANI doped with PSS has also been studied as an alternative to PSS: PEDOT. Recently, a composite of PANI/PSS: PEDOT/graphene has been fabricated, whose oxide has increased its conductivity, which shows great HTL in inverted PSCs by 10.5% power conversion efficiency (PCE). Therefore, a PANI composite doped from graphene nanomaterials, graphene oxide (GO) or reduced graphene oxide (RGO) can be a promising choice for HTL in inverse perovskites to achieve high efficiency. In general, PANIs can be an excellent polymer matrix to optimize the optoelectronic properties of HTLs in the structure of PSCs [43]. In addition, we have generally investigated the optoelectronic properties of PANI/graphene layers to investigate its capability as a new HTL in heterogeneous PSCs in an inverted model.

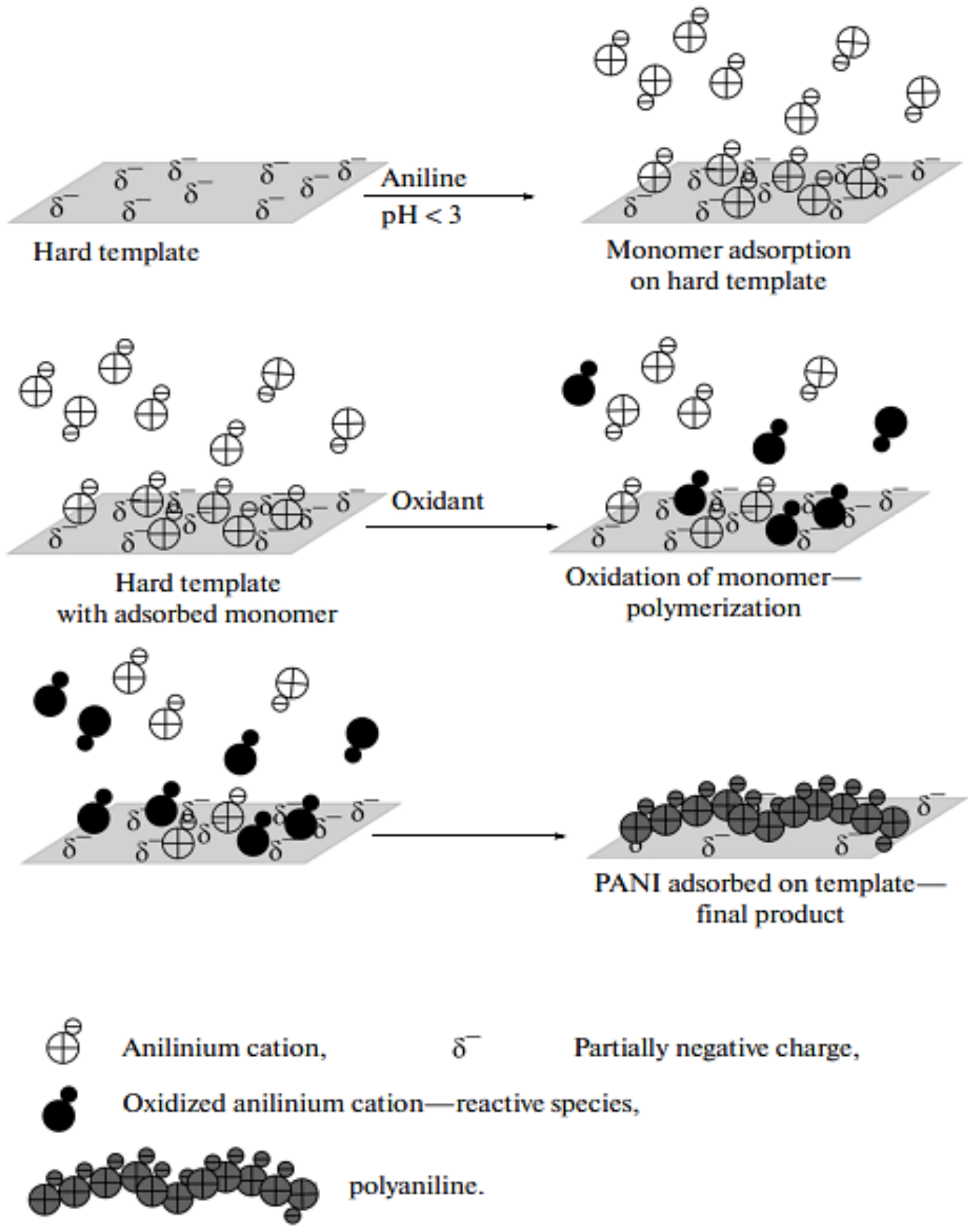

Figure 6. Neutral pH level and oxidation of polymer without PANI chain and oxidation with chain of PANI compounds

### 2.3. Materials and methods

To obtain this type of composite, the following method is used:

a) Poly (styrene sulfonic acid)-graft-polyaniline (PSSA-g-PANI) was synthesized.
b) The molar ratio of aniline to styrene sulfonic acid PSSA-g-PANI is set to 0.2% to optimize the electrical properties and improve its solubility in water.
c) GO is prepared using Hammers' improved method.
d) RGO was also used by p-toluene sulfonyl hydrazine as a reducing agent.
e) PbI2 (99.998%), Methyl ammonium iodide, PC61BM (99%, Nano-C), bathocuproine (BCP, 99%), were prepared from a commercial source.
f) We also use glass coated with indium tin oxide (ITO) for the substrate [44]–[46].

## 3. RESULTS AND DISCUSSION

In Figure 7, you can see the chemical structure of the material used in this method. PSSA-g-PANI is converted into a PANI copolymer that is in water and doped with itself, which forms the PSSA link. As a result, PSSA-g-PANI has higher conductivity and lower performance compared to the old PEDOT: PSS structure. For this reason, it can be considered as a suitable candidate. In order to obtain the results and optoelectronic properties of PANI/graphene nanocomposites, in this study, the kinds of graphene (GO and RGO) were mixed with PSSA-g-PANI. Two solutions of PSSA-g-PANI, referred to as PSSA-g-PANI+GO and PSSA-g-PANI+RGO are well dissolved in water, and the solution remains stable for several weeks, see Figure 7 [47]–[51]. In Figure 7, we see the different types of composite compositions in order. We see the composition of GO and RGO, and we see the effect of this type of reaction in polymer chain compounds.

The properties of perovskite absorbers, such as surface coverage, roughness, and crystallinity, are vital for optimizing perovskite properties. The conductivity is other important matter in perovskite properties. Because the optical current of PSCs depends on the electrical conductivity of the surface layers. Figure 8 shows the comparison of simulation results in PSCs. Transmittance spectra is shown in Figure 8(a), UPS kinetic energies in Figure 8(b), energy level diagram in Figure 8(c), and electrical conductivities of the films in Figure 8(d). As you can see in Figure 8(d), the PANI+GO film has better conductivity than the pristine PANI layer, and the PANI+RGO film layer shows almost higher conductivity than PANI. The high conductivity of PANI+graphene composite helps to increase hole transport and reduce recombination shown in Figure 8 [52]–[55]. As shown in Figure 8(d), the conductivity of the different types of composites can be compared in the standard state and in the two oxidized states, respectively, and it can be seen that the conductivity in the RGO state is much higher than in the standard state and the GO state.

Comparing simulation results of X-ray diffraction (XRD) analysis, UV-Vis absorption spectra, and steady state PL emission of the CH3NH3PbI3 layer on the composite HTLs are shown in Figure 9. The film crystallinity of the PSSA-g-PANI and the GO/RGO composites was obtained using XRD, shown in Figure 9(a). The XRD schemas of CH3NH3PbI3 films grown on three HTLs showed the tetragonal for the model. The absorption properties of the films were studied by UV-Vis absorption spectra, displaying that perovskite films were formed on the HTL films reckless of their composition in Figure 9(b). Also, Figure 9(c) shows the PL emission spectra of the $CH_3NH_3PbI_3$ films.

The surface morphology of the perovskite absorbers grown on different HTLs was obtained using FE-SEM, as shown in Figure 10. We observed smooth and homogeneous morphology with a comparable grain size of (250–400 nm) in the perovskite absorber surfaces reckless of the compositions of PSSA-g-PANI+RGO in Figure 10(a), PSSA-g-PANI+GO in Figure 10(b), PSSA-g-PANI in Figure 10(c), and PANI in Figure 10(d). In the FE-SEM image, the scale bar indicates 500 nm.

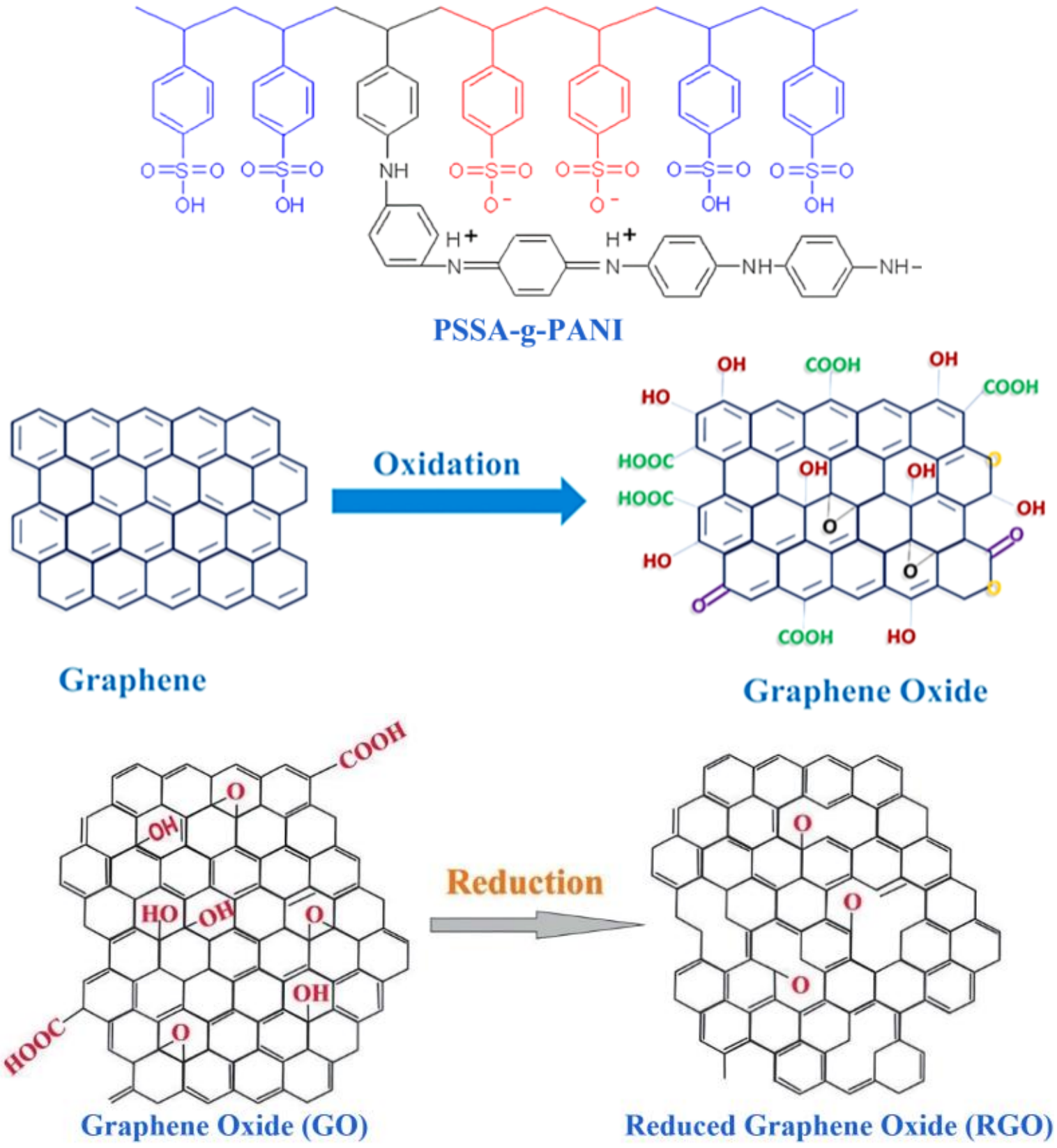

Figure 7. Molecular structure and aqueous solution of the materials in this study

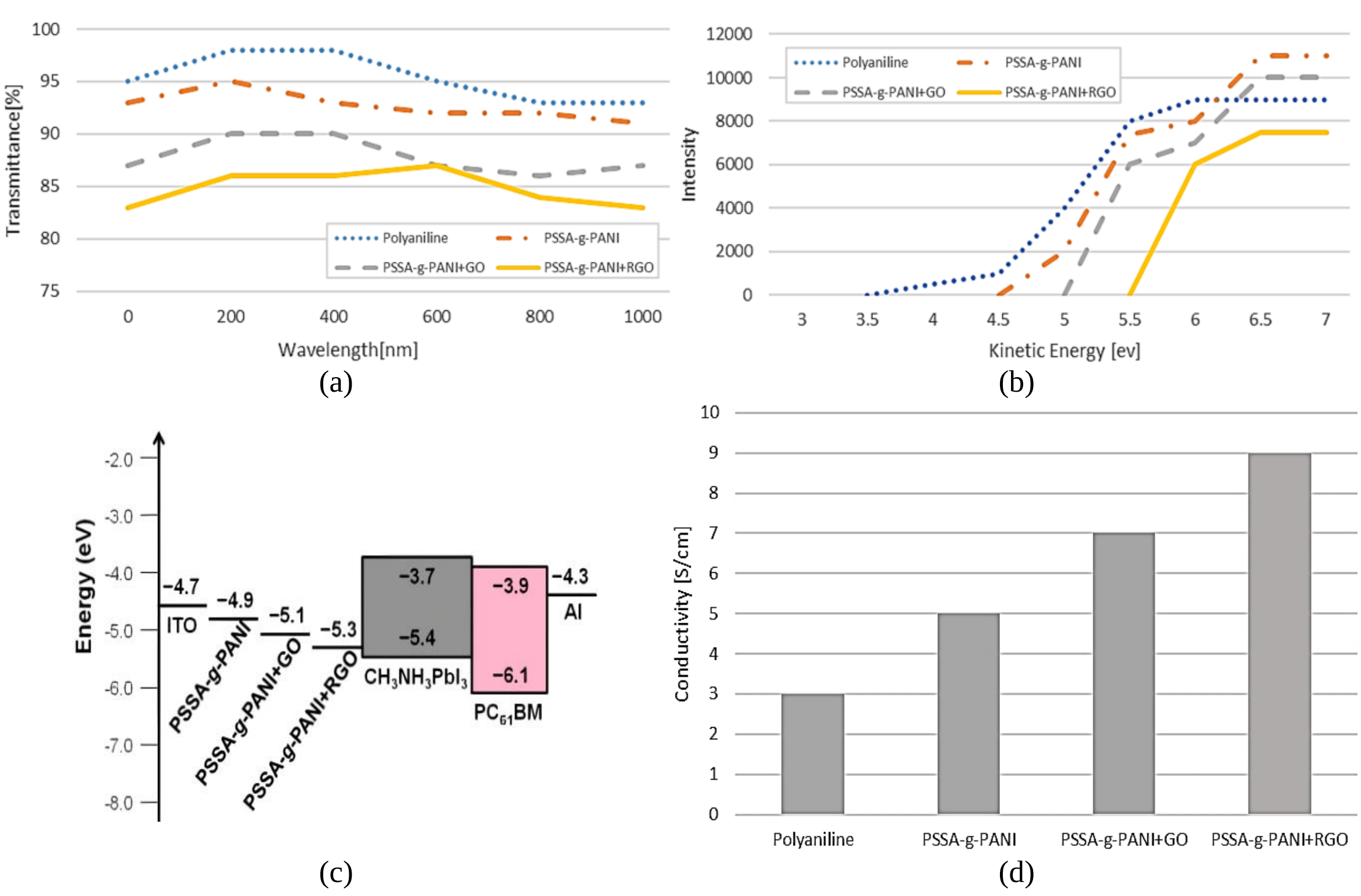

Figure 8. Comparing simulation results in PSCs: (a) transmittance spectra, (b) UPS kinetic energies, (c) energy diagram, and (d) conductivity of the composite films

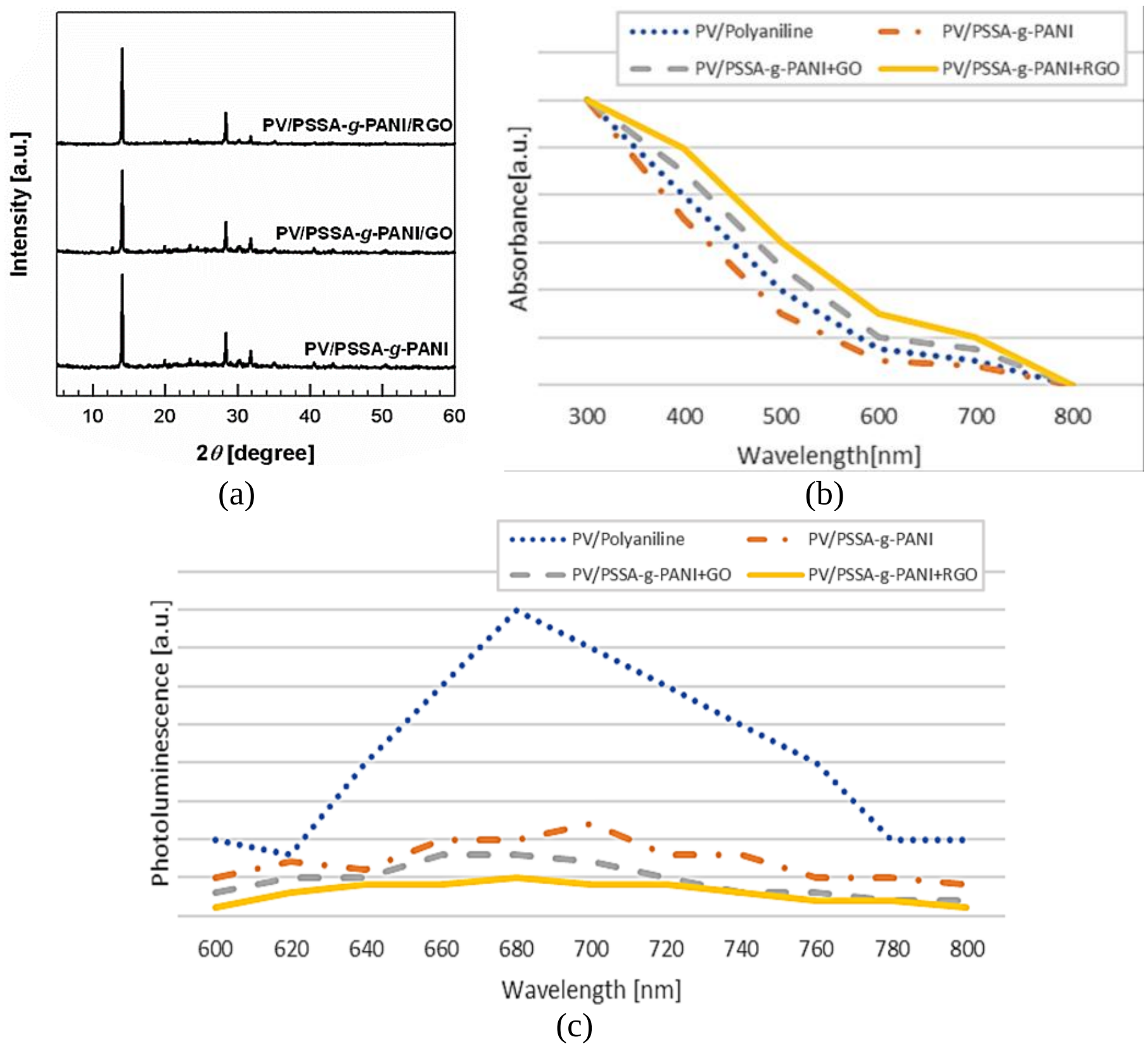

Figure 9. Comparing simulation results of (a) X-ray diffraction (XRD) analysis, (b) UV-Vis absorption spectra, and (c) steady state PL emission of the CH3NH3PbI3 layer on the composite HTLs

Figure 10. FE-SEM images of (a) PSSA-g-PANI+RGO, (b) PSSA-g-PANI+GO, (c) PSSA-g-PANI, and (d) PANI

Comparing simulation results of the J-V curves of the solar cells is displayed in Figure 11(a), while the external quantum efficiency (EQE) spectra are exhibited in Figure 11(b), and the decrease in power conversion efficiency (PCE) versus time (h) is shown in Figure 11(c). EQE for the sample with PANI+RGO shows the most power conversion potential at 400-700 nm, which shows the better efficiency of PANI/RGO among all PANI polymer-based HTLs. But PANI/RGO HTL has presented the lowest amount of hysteresis (HI) (Table 1), which shows the ability of PANI/RGO compared to other HTLs. In Table 1, according to the statistics that the researchers of this review article came across, it is more than 15 devices in each HTL [54], [55]. As can be seen in Table 1, the amount of open circuit voltage as well as the amount of short circuit current of all standard models and GO and RGO are compared. We can also see that the efficiency of the RGO mode is higher than the previous models.

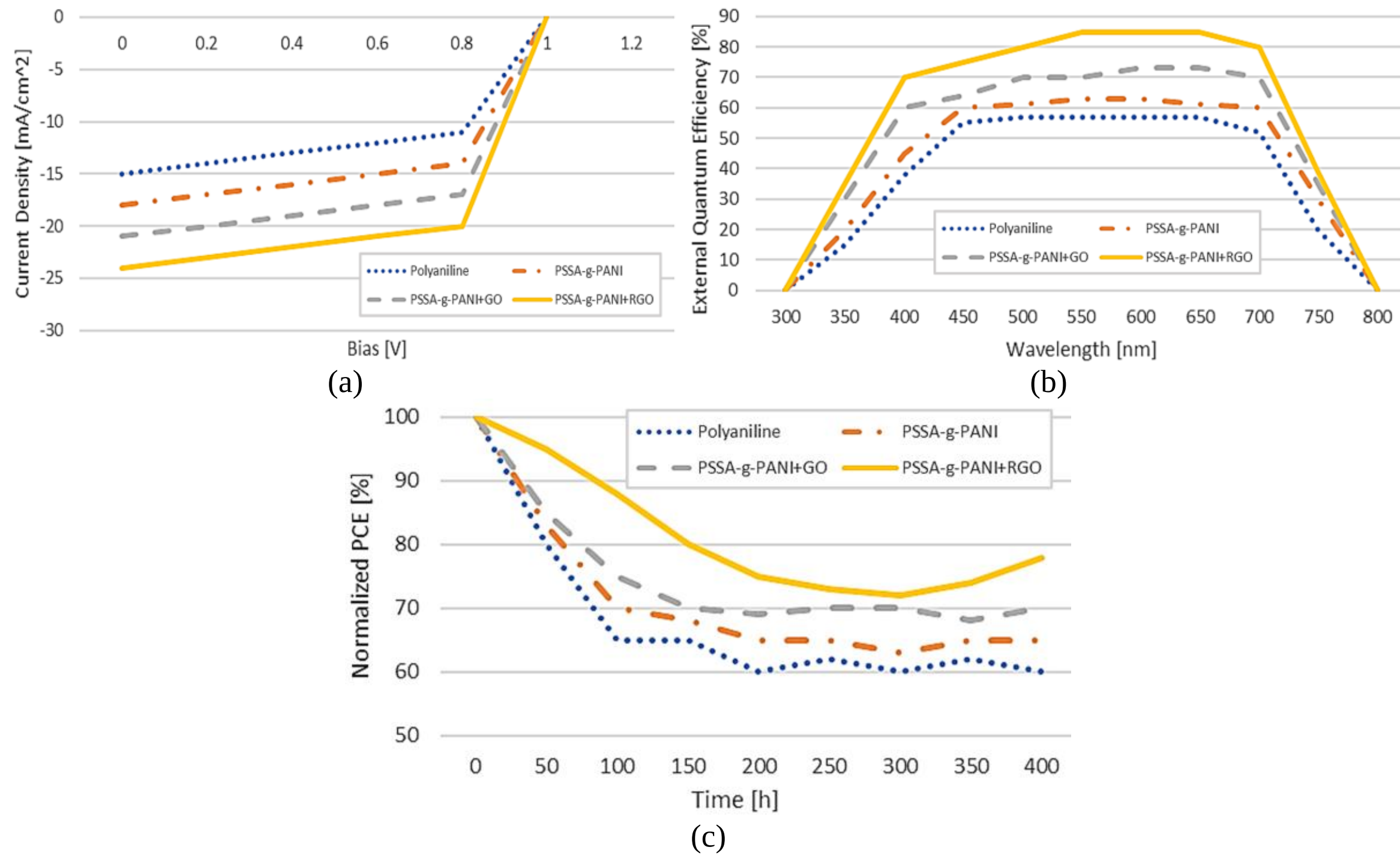

Figure 11. Comparing simulation results of (a) J-V characteristics (b) EQE spectra, and (c) PCE decrease versus time for composite HTLs

Table 1. The parameters of perovskite samples with various HTL films under 1 sun illumination

| HTL | Voc [v] | Jsc (mA/$cm^2$) | FF | PCE [%] | Rs [Ω] |
|---|---|---|---|---|---|
| Polyaniline | 0.88 | 18.2 | 0.51 | 12.2 | 3.01 |
| PSSA-g-PANI | 0.99 | 19.80 | 0.66 | 13.77 | 2.53 |
| PSSA-g-PANI+GO | 1.01 | 20 | 0.67 | 14.34 | 1.62 |
| PSSA-g-PANI+RGO | 1.05 | 22.42 | 0.72 | 16.70 | 0.72 |

## 4. CONCLUSION

In the structures of nano-solar cells, especially perovskite, productivity and stability are the most important elements in the manufacturing and innovation process. Every day, scientists try to increase these parameters by experimenting on new materials and using organic chemical compounds or polymer compounds. In this research, we continued our research by using the research of other researchers on a type of GO, which is used in a standard and reduced form, and the number of changes in the important parameters of this kind of solar cell, such as short circuit current and open circuit voltage, as well as We compared the amount of FF with each other and observed the effects of using graphite composites in the structure of solar cell layers. According to our observations in this article, the productivity increased from 12.2% to 13.77% and 14.34% and up to 16.7% respectively, which is a significant achievement. Also, the amount of FF after using RGO has reached 0.74. These results show us that the use of conductive polymers in the process of making perovskite solar cell layers, while saving the cost of making it, can be used to increase its efficiency.

## ACKNOWLEDGEMENTS

The authors gratefully acknowledge the University of Mazandaran (Babolsar, Iran), Islamic Azad University (Tehran, Iran), and Universitas Ahmad Dahlan (Yogyakarta, Indonesia) for providing the facilities essential to this research. Special thanks are also extended to the Embedded System and Power Electronics Research Group (ESPERG), Yogyakarta, for their invaluable support throughout this collaboration.

## FUNDING INFORMATION

The authors gratefully acknowledge funding from the University of Mazandaran and Islamic Azad University, Iran. We also appreciate the supplementary support offered by the Institute of Advanced Engineering and Science (IAES).

## AUTHOR CONTRIBUTIONS STATEMENT

This journal uses the Contributor Roles Taxonomy (CRediT) to recognize individual author contributions, reduce authorship disputes, and facilitate collaboration.

| Name of Author | C | M | So | Va | Fo | I | R | D | O | E | Vi | Su | P | Fu |
|---|---|---|---|---|---|---|---|---|---|---|---|---|---|---|
| Vahdat Nazerian | ✓ | ✓ | ✓ | ✓ | ✓ | ✓ | | ✓ | ✓ | ✓ | ✓ | | ✓ | |
| Mehran Hosseinzadeh Dizaj | ✓ | ✓ | | ✓ | ✓ | | ✓ | ✓ | ✓ | ✓ | | ✓ | | ✓ |
| Tole Sutikno | | ✓ | | ✓ | ✓ | | | | | ✓ | | ✓ | | ✓ |

C : **C**onceptualization
M : **M**ethodology
So : **So**ftware
Va : **Va**lidation
Fo : **Fo**rmal analysis
I : **I**nvestigation
R : **R**esources
D : **D**ata Curation
O : Writing - **O**riginal Draft
E : Writing - Review & **E**diting
Vi : **Vi**sualization
Su : **Su**pervision
P : **P**roject administration
Fu : **Fu**nding acquisition

## CONFLICT OF INTEREST STATEMENT

Authors state no conflict of interest.

## DATA AVAILABILITY

Data availability is not applicable to this paper as no new data were created or analyzed in this study.

## BIOGRAPHIES OF AUTHORS

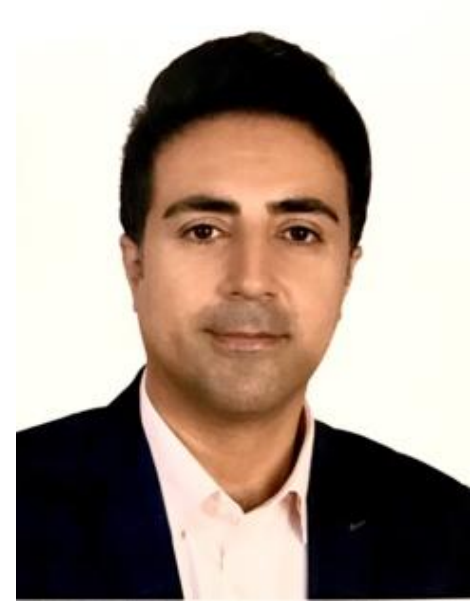

**Vahdat Nazerian** received his B.Sc. degree from Iran University of Science and Technology, Tehran, Iran, in 2003, M.Sc. and Ph.D. degrees from Khaje Nasir Toosi University of Technology, Tehran, Iran, in 2005 and 2011, respectively, all in Electronics. He joined the Electrical Engineering Department of the University of Mazandaran in 2014 as an assistant professor. His research and teaching concern semiconductor devices, analysis, and fabrication. His research presently focuses on soft computing, intelligent networks, nanotechnology and applications, renewable energy, and optimization algorithms. He can be contacted at email: v.nazerian@umz.ac.ir.

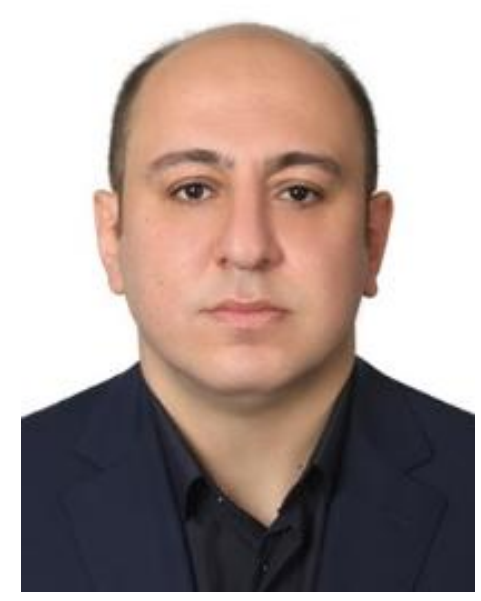

**Mehran Hosseinzadeh Dizaj** received his B.Sc. degree in electrical-control engineering (precision instruments) and M.Sc. degree, and Ph.D. degree in electrical and electronic engineering (Nano and microelectronics). The field of activity is about semiconductor devices and perovskite solar cells. Currently, he is engaged in teaching and laboratory activities in the nanophysics and thin film laboratory at Azad University of Tehran, Central and Tehran East. He can be contacted at email: meh.hosseinzadehdizaj@iauctb.ac.ir.

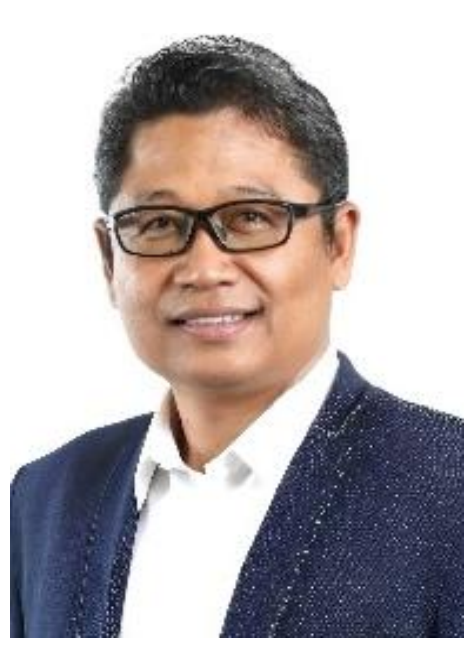

**Prof. Ir. Tole Sutikno, Ph.D., MIET, IPM., ASEAN Eng.** is a full professor in the Department of Electrical Engineering at Universitas Ahmad Dahlan (UAD) in Yogyakarta, Indonesia. He has held this position since 2023, having previously served as an associate professor from 2008. He earned his bachelor’s degree from Universitas Diponegoro in 1999, his master’s degree from Universitas Gadjah Mada in 2004, and his Ph.D. in Electrical Engineering from Universiti Teknologi Malaysia in 2016, where his doctoral research focused on advanced digital power electronics and intelligent control systems. From 2016 to 2021, he served as the Director of the Institute for Scientific Publishing and Publications (LPPI) at UAD, where he led initiatives to strengthen research visibility, journal management, and international collaboration in scholarly publishing. Since 2024, he has served as the head of the master's program in electrical engineering at UAD, following his leadership of the undergraduate program in electrical engineering in 2022. He is also the founding leader of the Embedded Systems and Power Electronics Research Group (ESPERG), which actively collaborates with both national and international institutions on topics such as fault-tolerant embedded systems, FPGA-based control, and renewable energy integration. He is widely acknowledged for his contributions to digital design, industrial electronics, motor drives, robotics, intelligent systems, and AI-based automation. His interdisciplinary research emphasizes practical deployments in industrial and healthcare contexts, covering FPGA applications, embedded systems, power electronics, and digital libraries. He has published 380 peer-reviewed articles in high-impact journals and conferences indexed by Scopus. As of 2025, Google Scholar indicates over 6,000 citations, with an h-index of 36 and an i10-index of 174. In recognition of his global research impact, he has been listed among the Top 2% of Scientists Worldwide by Stanford University and Elsevier BV from 2021 to the present, a distinction based on standardized citation metrics across all scientific disciplines. He can be contacted at email: tole@te.uad.ac.id.